\documentclass[apj,iop,twocolumn]{emulateapj}
\usepackage[dvipdfm]{hyperref}
\submitted{accepted for publication in The Astrophysical Journal}

\newcommand {\ea} {{\it et~al.}}
\newcommand {\be} {\begin{equation}}
\newcommand {\ee} {\end{equation}}

\shorttitle{$\gamma$-ray/optical lags in blazars}
\shortauthors{Janiak \ea}

\begin{document}
\title{On the origin of the $\gamma$-ray/optical lags in luminous blazars}

\author{Mateusz~Janiak\altaffilmark{1},
Marek~Sikora\altaffilmark{1},
Krzysztof~Nalewajko\altaffilmark{2},
Rafa{\l}~Moderski\altaffilmark{1}, 
and Greg~M.~Madejski\altaffilmark{3}}

\altaffiltext{1}{Nicolaus Copernicus Astronomical Center, Bartycka 18,
  00-716 Warsaw, Poland; {\tt mjaniak@camk.edu.pl}, {\tt sikora@camk.edu.pl}}
\altaffiltext{2}{University of Colorado, UCB 440, Boulder, CO 80309,
  USA}
\altaffiltext{3}{Kavli Institute for Particle Astrophysics and
  Cosmology, SLAC National Accelerator Laboratory, Stanford
  University, 2575 Sand Hill Road M/S 29, Menlo Park, CA 94025, USA}

\begin{abstract}
Blazars are strongly variable sources that occasionally show spectacular 
flares visible in various energy bands.  These 
flares are often, but not always, correlated.  In a number of cases
the peaks of optical flares are found to be somewhat delayed with
respect to the gamma-ray peaks.  One notable example of such a delay
was found in 3C~279 by Hayashida et~al. and interpreted as a result
of steeper drop with distance of the energy density
of external radiation
field than of the magnetic energy density.  In this paper we
demonstrate that in general, depending on the respective energy
density profile along the jet, such lags 
can have both signs and that they can take place for any
ratio of these energy densities.  We study the dependence of such lags
on the ratio of these energy densities at a distance of a maximal 
energy dissipation in a jet, on their gradients, as well as on the
time profile of the relativistic electron injection within the moving
source.  We show how prominent such lags can be, and what are their
expected time scales.  We suggest that studies of such lags can
provide a powerful tool to resolve the structure of relativistic jets
and their radiative environment.  As an example we model the lag
observed in 3C~279, showing that in this object the flare is produced
at a distance of a few parsecs from the central black hole, consistent
with our previous inferences based on the spectra and optical
polarization properties.
\end{abstract}


\section{Introduction}

Blazars are variable in all spectral bands and over a very broad range
of time scales, from years down to days or sometimes even minutes
\citep[e.g.,][]{U97, Aa10a, Gh08}.  Largest amplitude variations 
are detected in
the gamma-ray band \citep{Aa10a}.  In luminous blazars, belonging to
the class of Flat-Spectrum Radio Quasars (FSRQs), the gamma-ray light
curves usually correlate with the optical light curves \citep{C12}, but
with the peaks of short-term flares sometimes delayed with respect to
each other \citep{W09}.  In particular, gamma-ray flares leading the
optical flares by a few days have been indicated in several blazars ---
PKS~1510-089 \citep{Aa10b,DA11}, PKS~1502+106 \citep{Aa10c},
AO~0235+164 \citep{Ag11,Ac12}, and 3C~279 \citep{Ha12}.

Such lags are very intriguing because, according to the most commonly accepted 
external-radiation-Compton (ERC) models, radiation in the optical and
$\gamma$-ray bands is produced by electrons with similar energies and
deeply in the fast cooling regime (i.e.\ electron cooling time scale
is much shorter than the source light-crossing time scale).  In such
case, lags cannot result from the difference in cooling time scales,
as in the scenarios considered by \citet{2001ApJ...554....1S} and
\citet{SM05}.  \citet{Ha12} suggested that the optical/$\gamma$-ray
lags may result from different distance dependencies of the magnetic
energy density and the energy density of external diffuse radiation
field.  Here we explore this possibility in detail, studying the 
dependence of the lag properties on such model parameters as gradients
of the energy densities, the electron injection time profile, and the
ERC-to-synchrotron flux ratio at the distance of the maximum electron
injection rate.

Our paper is structured as follows. In Section \ref{sec_model}, 
we describe our model of synchrotron/ERC flares that allows us 
to study the $\gamma$-ray/optical lags, and study how the resulting 
lag parameters depend on the details of the model. In Section \ref{sec_appl}, 
we apply our model to luminous blazars with dense radiative environment, 
and to the particular case of 3C~279. A further discussion and 
conclusions are given in Section \ref{sec_disc}.

\section{Modeling the lags}
\label{sec_model}

\subsection{Assumptions and equations}

Assuming that the source of a non-thermal flare propagates down the
jet with a constant velocity, we can describe its activity as a
function of its distance from the black hole, $r = c t' \Gamma$, where
$t'$ is the time measured in the source co-moving frame and $\Gamma$
is the bulk Lorentz factor of the jet. We investigate light curves
produced by electrons deeply in the fast cooling regime by the
synchrotron and ERC processes.
They are determined assuming that the observed time scale of light-travel 
effects caused by the finite size of a source is much shorter than the observed
time scale of the source propagation. Such assumption is justified by
observations suggesting that the AGN jet semi-opening angles are smaller than 
the Doppler angles (e.g. \cite{Pu09}). 
The ERC process is treated in the
Thomson regime, using the following approximate formula
\citep{2003A&A...406..855M}:
\be [\nu F_{\rm ERC,\nu}]_{\nu=\nu_{\rm HE}}(t_{\rm obs}) \propto
    [\gamma N_{\gamma}|\dot \gamma|_{\rm ERC}](r)\,,
\label{eq:Ferc}
\ee
for the electron Lorentz factor satisfying
$\gamma^2 \simeq (1+z)\nu_{\rm HE}/({\cal D}^2\nu_{\rm ext})$,
where $\nu_{\rm HE}$ is the
observed frequency of the high-energy component, $\nu_{\rm ext}$ is
the average frequency of the external diffuse radiation field,
$N_{\gamma}$ is the electron energy spectrum, $|\dot \gamma|_{\rm ERC}
\propto \gamma^2 u_{\rm ext}'(r)$ is the rate of the electron energy
losses via the ERC process, $u_{\rm ext}'(r)$ is the energy density of
external radiation field as measured in the source comoving frame,
$t_{\rm obs}=r/({\cal D}\Gamma c)$ is the observation time, ${\cal
  D}=1/[\Gamma(1-v\cos{\theta_{\rm obs}/c})]$ is the Doppler
factor, and $z$ is the source redshift.

The synchrotron flux is approximated in an analogous way by
\be
[\nu F_{{\rm syn},\nu}]_{\nu=\nu_{\rm LE}}(t_{\rm obs}) \propto
    [\gamma N_{\gamma}|\dot \gamma|_{\rm syn}](r)\, ,
\label{eq:Fsyn}
\ee
for the electron Lorentz factors $\gamma^2 \simeq 3\pi(1+z){\rm m_e
  c}\nu_{\rm syn}/[2 {\rm e}B'(r)]$, where $\nu_{\rm LE}$ is the
observed frequency of the low-energy spectral component,
$|\dot\gamma|_{\rm syn} \propto \gamma^2 u_B'(r)$ is the rate of the
electron energy losses via the synchrotron process, and
$u_B'(r)=B'(r)^2/(8\pi)$ is the energy density of the magnetic field
carried by the source.  Assuming a power-law injection function
$Q_{\gamma} = K(r) \gamma^{-p}$, where $K(r)$ has a maximum at
$r=r_{\rm m}$, one can find that deeply in the electron fast cooling
regime
\be
N_{\gamma} = {\int_{\gamma} Q_{\gamma} d\gamma \over |\dot
  \gamma|_{\rm rad}} \propto {\int_{\gamma} Q_{\gamma} d\gamma \over
  \gamma^2 (u_B' + u_{\rm ext}') } \propto {K(r) \gamma^{-p-1} \over (u_B'
  + u_{\rm ext}')} \,,
\label{eq:Ng}
\ee
where we ignore radiative losses due to the synchrotron self-Compton
(SSC) process.

Combination of (\ref{eq:Ng}), (\ref{eq:Ferc}), and (\ref{eq:Fsyn})
gives
\be [\nu F_{\rm ERC,\nu}]_{\nu=\nu_{\rm HE}} \propto \gamma^{-p+2} K(r) 
{u_{\rm ext}'(r) \over u_{\rm ext}'(r) + u_{\rm B}'(r)} \,,\ee
\be [\nu F_{\rm syn,\nu}]_{\nu=\nu_{\rm LE}} \propto \gamma^{-p+2} K(r)
{u_{B}'(r) \over u_{\rm ext}'(r) + u_B'(r)}\,.\ee

For particular spatial distributions of the energy densities $u_{\rm
  ext}'(r) \propto r^{-\beta_E}$ and $u_{\rm B}'(r) \propto
r^{-\beta_B}$, we obtain that the ERC and synchrotron light curves are
\be F_{ERC, \nu_{HE}}(x) \propto K(x) 
{q_m \over q_m + x^{\Delta \beta}} \,,\label{eq:Lerc_ana}\ee
and
\be F_{syn,\nu_{LE}}(x) \propto x^{-(p-2)\beta_B/4} \, K(x) {x^{\Delta
    \beta} \over q_m + x^{\Delta \beta}} \,,\label{eq:Lsyn_ana}\ee
where $q_m = u_{\rm ext}'(r_m) / u_B'(r_m)$, $x=r/r_m$, and $\Delta\beta
= \beta_E-\beta_B$.
\footnote{Note that the factor $x^{-(p-2)\beta_B/4}$ in the latter equation
results from the fact that in the synchrotron process 
$\gamma \propto 1/\sqrt{B(r)'} \propto r^{\beta_B/4}$.}

In order to compute the light curves, we need to specify the electron
injection function $K(x)$.  We choose a modified Gaussian function
\be K(x) = x \cdot \exp\left[-\frac{(x-1)(x+\sigma^2-1)}{\sigma^2}\right]
\label{eq:injection}\ee
with a `dispersion' $\sigma$ and the property that $K(x)\to 0$ for
$x\to 0$.

\subsection{Light curves}

Equations (\ref{eq:Lerc_ana}), (\ref{eq:Lsyn_ana}) and
(\ref{eq:injection}) show that the number of free parameters
determining the light curves is: 3 for the ERC case ($q_m$, $\Delta
\beta$ and $\sigma$); and 5 for the synchrotron case (the two
additional parameters are $\beta_B$ and $p$).  The number of free
parameters determining the synchrotron light curve can be reduced to 3
by assuming the electron injection function index of $p=2$.  Noting
that electrons injected with such an index in the fast cooling regime
produce spectra with the index $\alpha = p/2 = 1$ ($F_\nu \propto
\nu^{-\alpha}$), and that $\alpha \sim 1-1.4$ are typical slopes of
synchrotron spectra of luminous blazars \citep{lo85}, we consider this
choice to be representative, and, with the exception of Section
\ref{sec:pne2}, we limit our further investigations to the case of
$p=2$.

Before plotting some examples of the light curves and their dependence
on the model parameters, we can deduce the main features of the light
curves directly from Eqs~(\ref{eq:Lerc_ana}) and (\ref{eq:Lsyn_ana}).
We can see that:
\begin{enumerate}
\item
For $\Delta\beta=0$, the ERC and synchrotron light curves follow the
shape of the electron injection function and both peak at $x=1$ (zero
lag);

\item
From the ratio of the two light curves ($\propto x^{\Delta\beta}$) one
can deduce that for $\Delta\beta > 0$ the ERC peak precedes the
synchrotron one (hereafter we call it the positive lag), and for
$\Delta\beta < 0$ the ERC peak is delayed relative the synchrotron one
(negative lag);

\item
For $q_m \gg 1$, the shape of the ERC flare follows the shape of the
electron injection function, and therefore has a peak at $x\simeq 1$
\citep[it was the case studied in the Appendix to][]{Ha12}.  For $q_m
\ll 1$, the synchrotron flare follows the shape of the electron
injection function and has a peak at $x \simeq 1$.
\end{enumerate}

\begin{figure}
  \begin{center}
    \includegraphics[width=0.45\textwidth, bb=63 144 550 660, angle=-90]{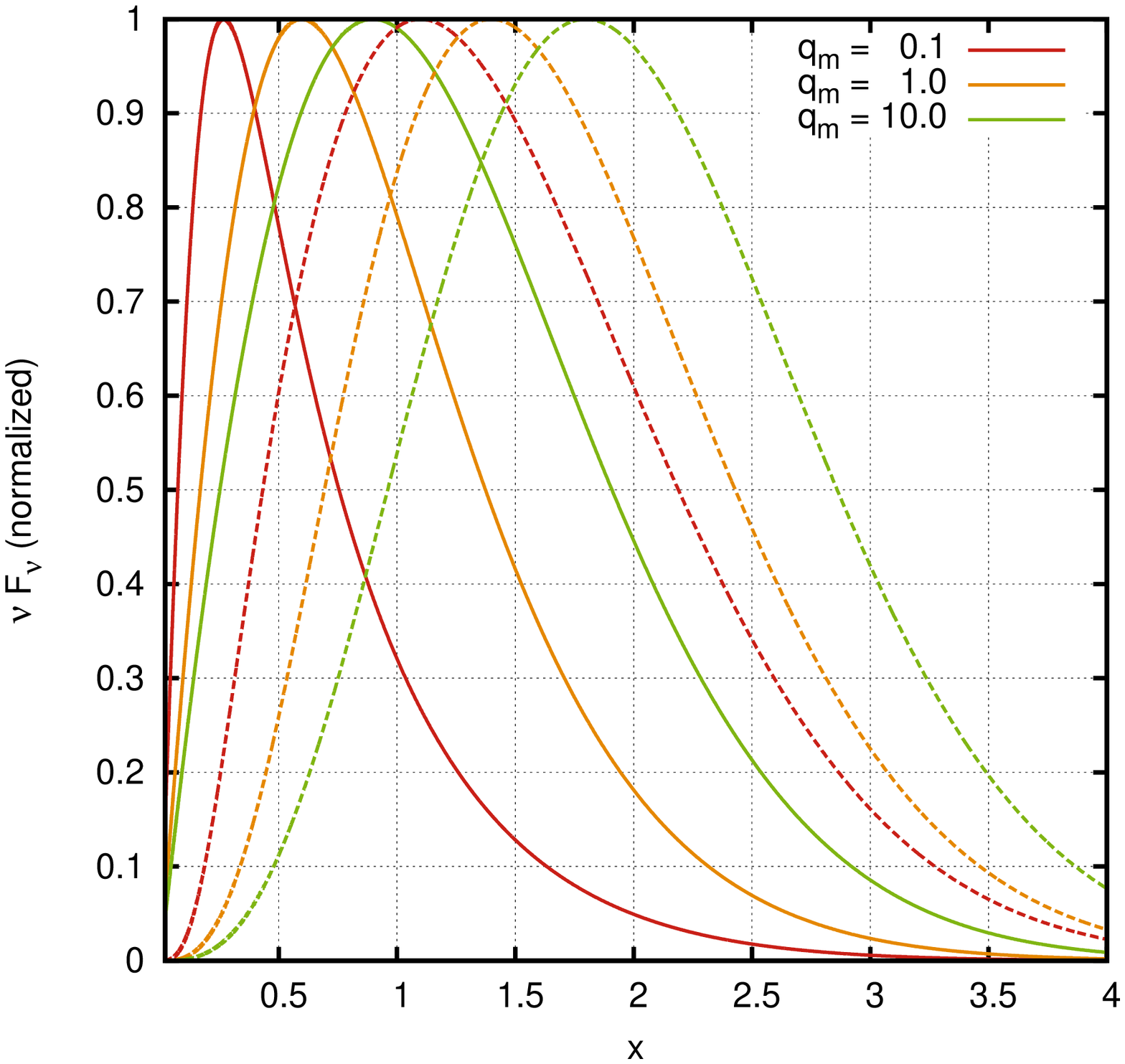}
    \includegraphics[width=0.45\textwidth, bb=63 144 550 660, angle=-90]{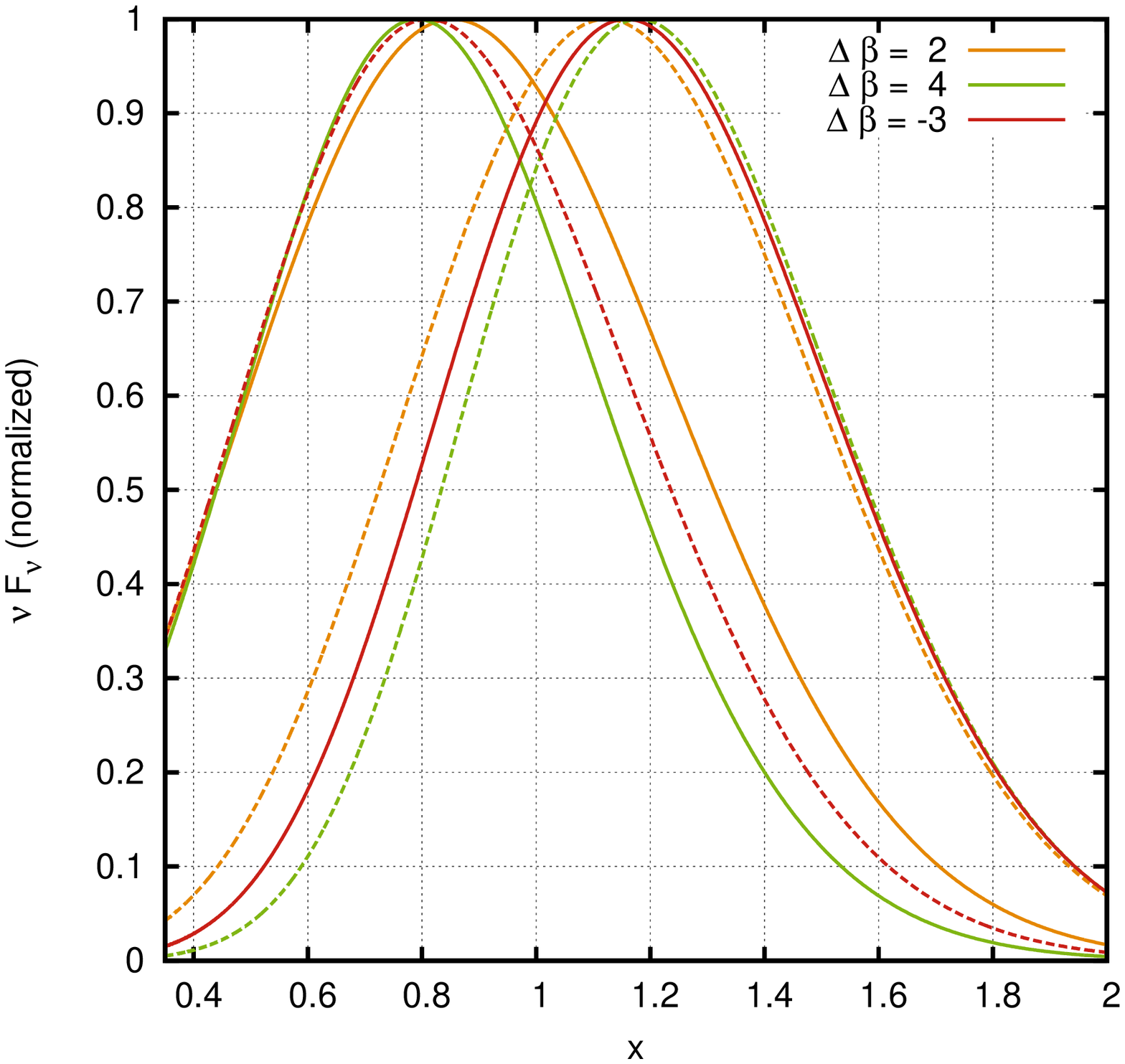}
    \includegraphics[width=0.45\textwidth, bb=63 144 550 660, angle=-90]{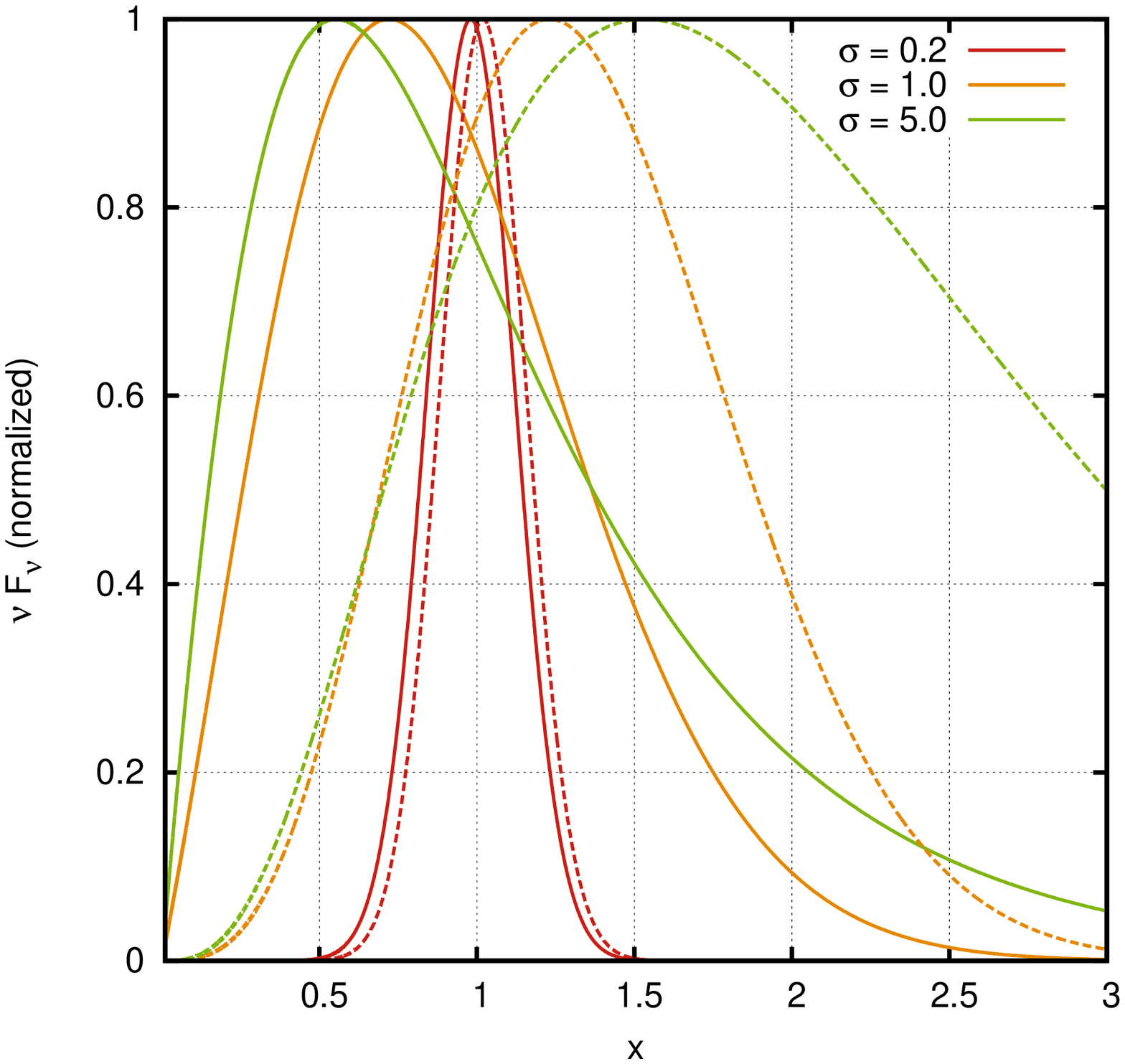}
    \caption{Normalized shapes of ERC flares (\emph{solid lines}) and
      synchrotron flares (\emph{dotted lines}).
      \emph{Top left panel:} $\sigma=0.6$ and $\Delta\beta=2$.
      \emph{Top right panel:} $\sigma=0.6$ and $q_{m}=1$.
      \emph{Bottom panel:} $\Delta\beta=2$ and $q_{m}=1$.}
    \label{fig:flare} 
  \end{center}
\end{figure}

In Figure~\ref{fig:flare}, we present three plots showing the dependence
of the light curves on $q_m$, $\Delta\beta$ and $\sigma$.  We can see
that $q_m$ determines mainly the location of the flare peaks relative
to the point $x=1$, while $\sigma$ and $\Delta\beta$ determine mainly
the distance between peaks and the flare symmetry.

\subsection{The length and prominence of a lag}

\begin{figure}
  \begin{center}
    \includegraphics[width=0.45\textwidth, bb=68 144 550 660, angle=-90]{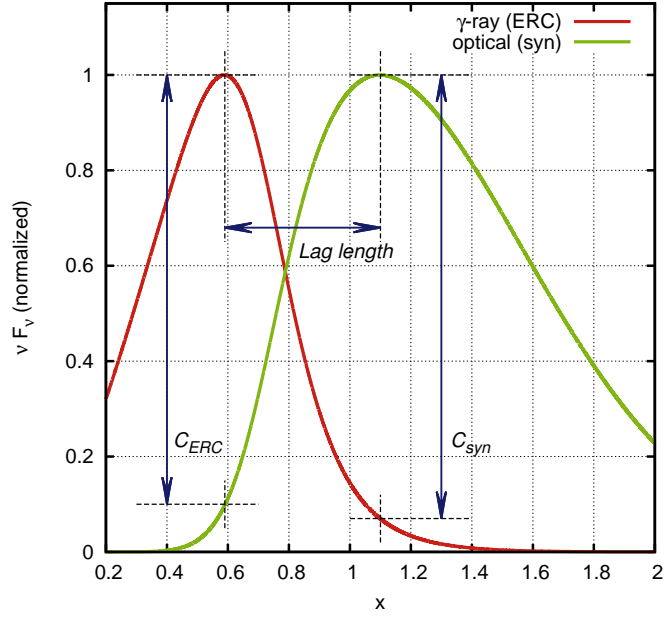}
    \caption{Definitions of the lag length and the two measures of lag
      prominence for a sample pair of normalized light curves.}
    \label{fig:def}
  \end{center}
\end{figure}

\begin{figure}
 \begin{center}
  \includegraphics[width=0.45\textwidth, bb=63 147 536 665, angle=-90]{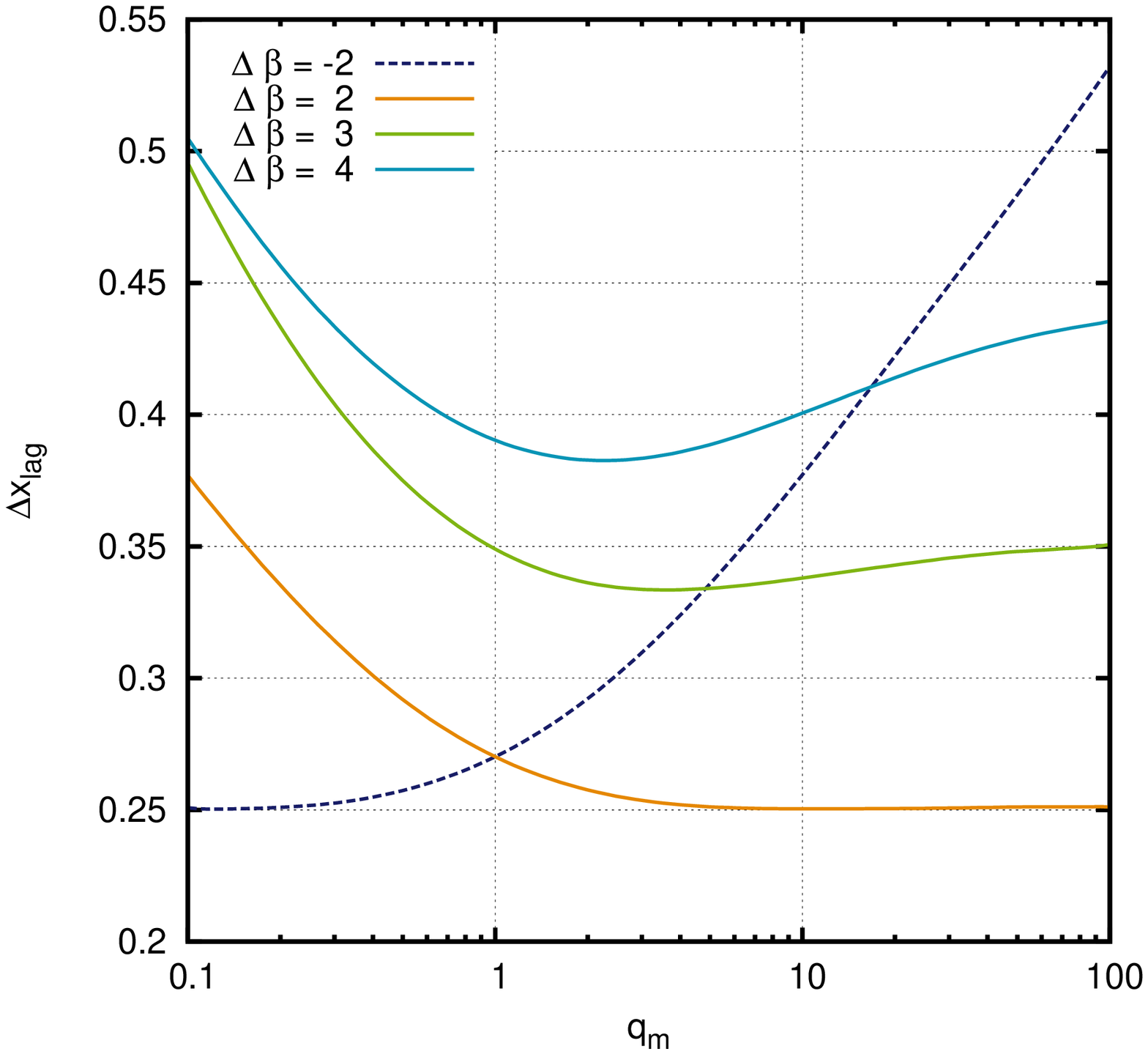}
  \includegraphics[width=0.45\textwidth, bb=63 147 536 665, angle=-90]{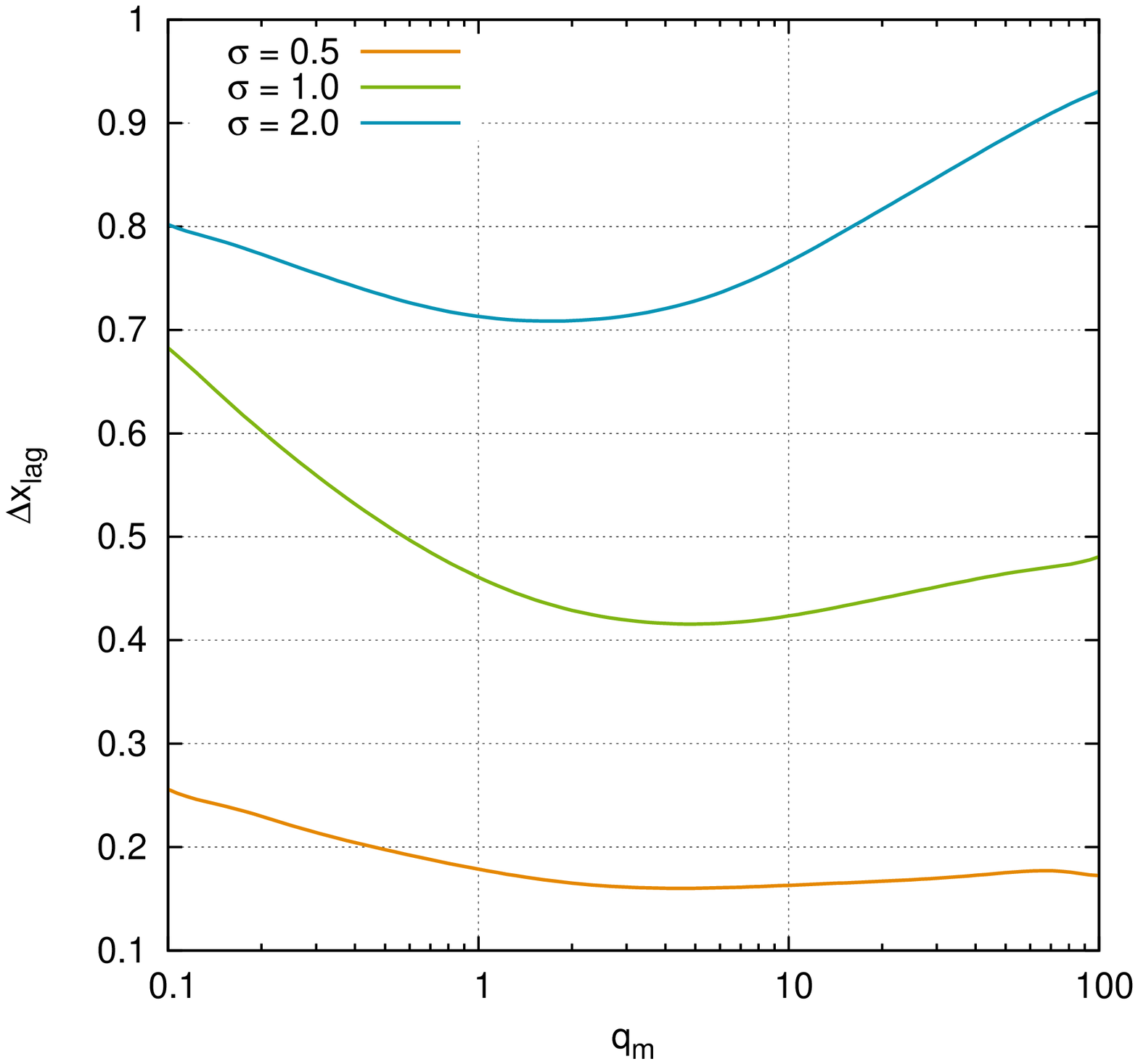}
  \caption{Lag length dependence on $q_{m}$.
    \emph{Left panel:} $\sigma=0.6$. Dotted curve, corresponding to $\Delta \beta<0$ (negative lag), has been flipped over x-axis.
    \emph{Right panel:} $\Delta\beta=2$.}
  \label{fig:lag} 
  \end{center}
\end{figure}

We define the lag length as
\be \Delta x_{\rm lag} = x_{\rm syn} - x_{\rm ERC} \,,\ee
(see Fig.~\ref{fig:def}) where $x_{\rm ERC}$ and $x_{\rm syn}$ 
are the locations where the ERC and
synchrotron peaks are produced.  The dependence of $\Delta x_{\rm
  lag}$ on $q_m$, $\Delta\beta$ and $\sigma$ is presented in
Fig.~\ref{fig:lag}.  One can see that for $\sigma=0.6$,
$\Delta\beta=4$ and $q_m >1$, the lag length reaches the value of
$\Delta x_{\rm lag} \sim 0.4$ and may be much higher for higher values
of $\sigma$.

\begin{figure}
 \begin{center}
  \includegraphics[width=0.45\textwidth, bb=68 154 536 661, angle=-90]{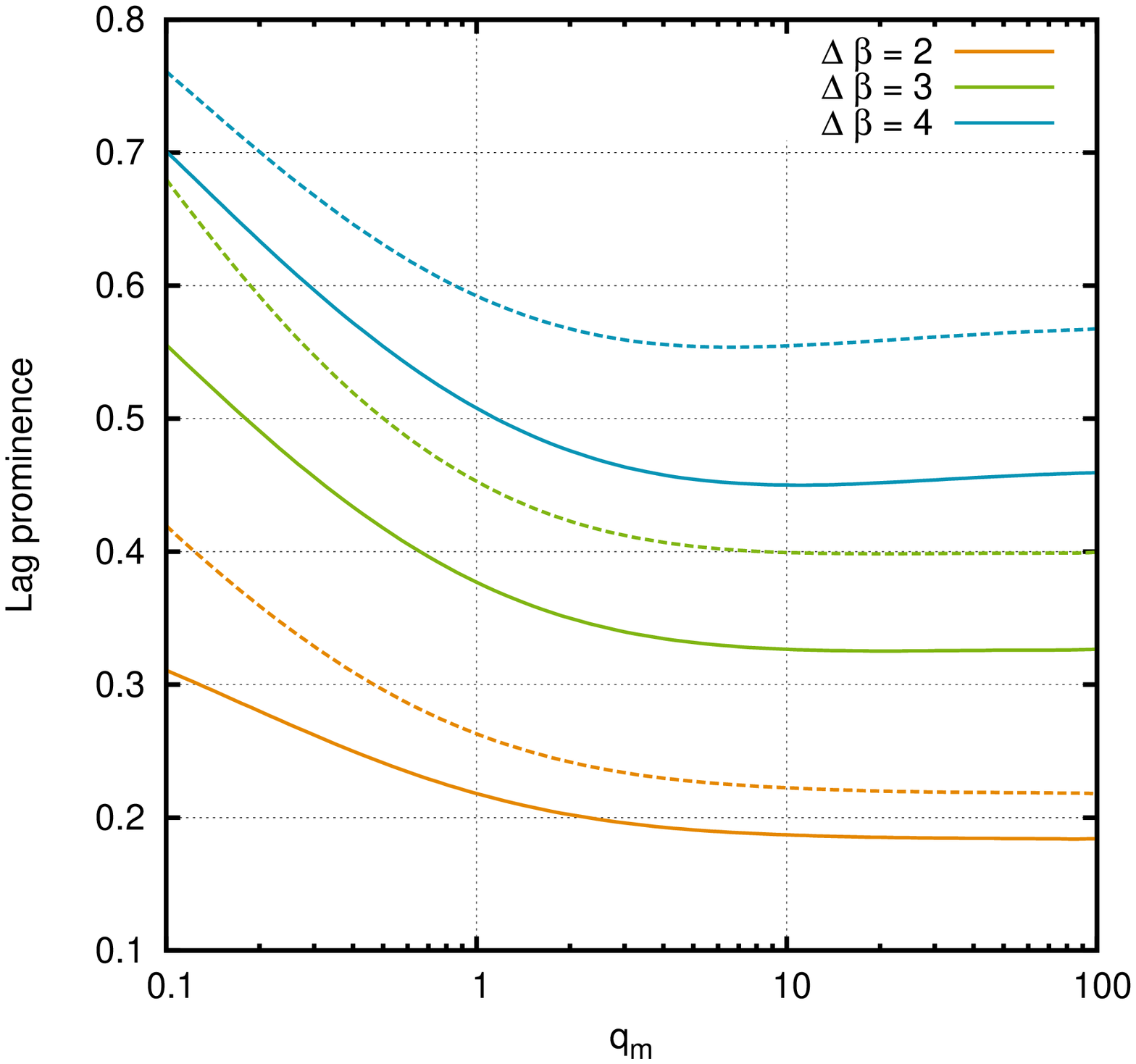}
  \includegraphics[width=0.45\textwidth, bb=68 154 536 661, angle=-90]{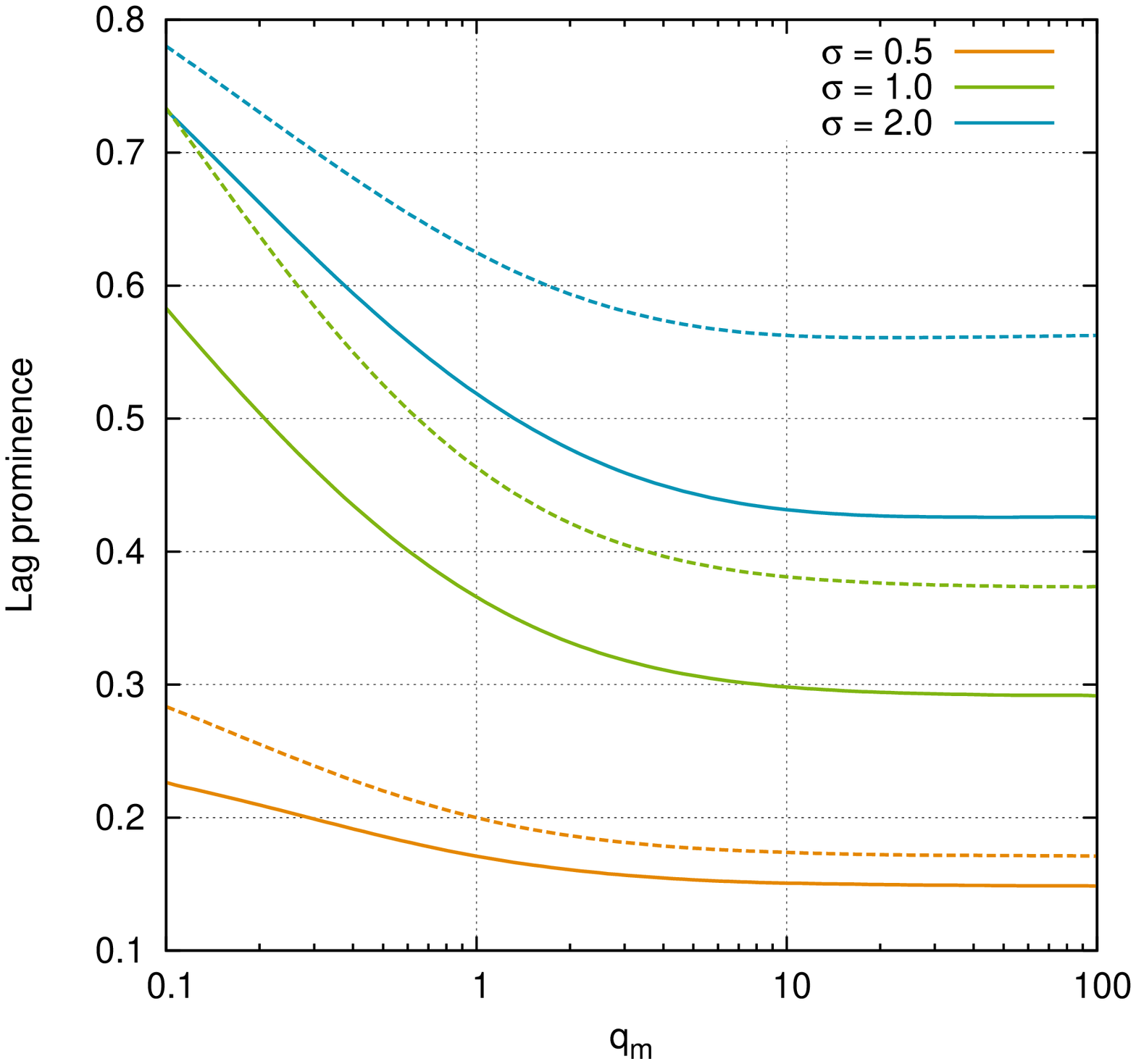}
    \caption{Lag prominence indicators, $C_{\rm ERC}$ (\emph{solid lines}) and
      $C_{\rm syn}$ (\emph{dotted lines}), dependence on $q_{m}$. \emph{Left panel:} $\sigma=0.6$.
    \emph{Right panel:} $\Delta\beta=2$.}
    \label{fig:contrast}
 \end{center}
\end{figure}

However, as one can see from Fig.~\ref{fig:flare}, even for very large
lag lengths the flares overlap significantly and are not strongly 
contrasted.  To verify for which model parameters are these lags
reasonably prominent, we introduce two parameters to quantify the lag
`prominence' (see Fig.~\ref{fig:def}):\footnote{Note that our choice of
  description of the lag prominence with two parameters is dictated by
  practical reasons.  Because of typically poor time sampling and
  large errors in high energy light curves, such definition enables us to
  estimate the lag contrast in a more precise way than using a single
  parameter, e.g.\ the flux at the intersection of two normalized
  flares.}
\be C_{\rm ERC} = {F_{\nu,{\rm HE}}(x_{\rm ERC}) - F_{\nu,{\rm HE}}(x_{\rm syn})
\over F_{\nu,{\rm HE}}(x_{\rm ERC})} \,\label{eq:prom_erc}\ee
and
\be C_{\rm syn} = {F_{\nu,{\rm LE}}(x_{\rm syn}) - F_{\nu,{\rm LE}}(x_{\rm ERC})
\over F_{\nu,{\rm LE}(\gamma)}(x_{\rm syn})} \,.\label{eq:prom_syn}\ee
The dependence of these indicators on the model parameters is
presented in Fig.~\ref{fig:contrast}.
We find that both measures of the lag prominence increase with 
increasing $\Delta\beta$ and increasing $\sigma$.
The dependence on $q_{\rm m}$ is more complex -- for $q_{\rm m}\lesssim 1$ 
the lag prominence decreases with increasing $q_{\rm m}$, while 
for $q_{\rm m}\gg 1$ the lag prominence depends on $q_{\rm m}$ very weakly.

\subsection{The case of $p\ne 2$}
\label{sec:pne2}
\begin{figure}
  \begin{center}
    \includegraphics[width=0.45\textwidth, bb=68 154 536 661, angle=-90]{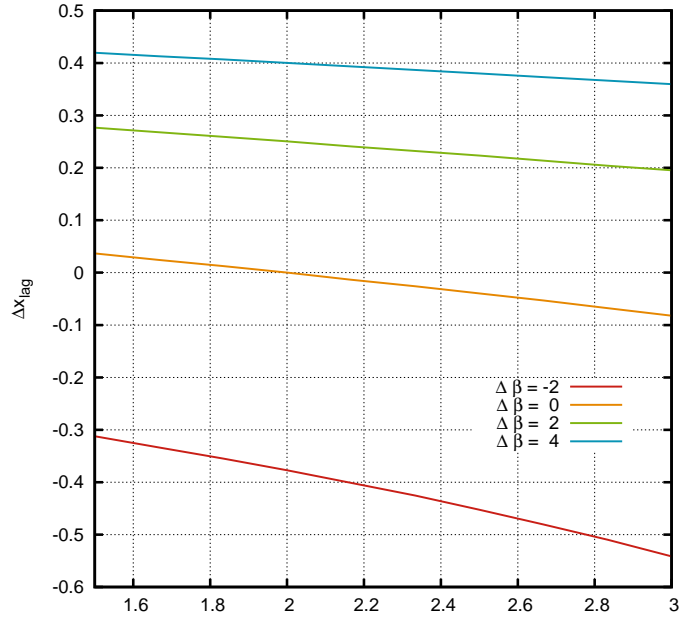}
    \caption{Lag length dependence on the index of the electron
      injection function $p$ for $q_m=10$, $\beta_B=2$, $\sigma=0.6$.}
    \label{fig:roznep}
  \end{center}
\end{figure}

In Figure~\ref{fig:roznep}, we show the dependence of the lag length
on the injected electron energy distribution index $p$. We can see
that regardless of the value of $\Delta\beta$, the lag length slightly
decreases with increasing $p$. This means that the absolute value of
the lag decreases for positive lags and increases for negative
lags. This can be easily understood from the existence of a factor
$x^{-(p-2)\beta_B/4}$ in Eq.~(\ref{eq:Lsyn_ana}).

\section{Applications}
\label{sec_appl}
We have studied basic features of lags between the ERC and synchrotron
light curves produced in the fast cooling regime. 
Our model shows that the presence of
such lags is made possible by a non-zero value of 
$\Delta \beta$ which corresponds to a significant 
change of the Compton dominance parameter 
$q=u_{\rm ext}'/u_B' \propto r^{-\Delta \beta}$ within the distance range of
enhanced activity of the propagating source.  In a quasar radiative
environment the change of $q$ with distance is expected to result mainly
from the strong stratification of the external radiation field
sources, being connected with broad-line region (BLR) and hot-dust
region (HDR).  Changes of $q$ with distance can be deduced from
Fig.~\ref{fig:ur}, where radial distributions of respective energy
densities are schematically presented \citep[see
  also][]{2009ApJ...704...38S}.  In this figure we marked regions
where $q$ decreases with a distance (region ``I'') and regions where
$q$ increases with a distance (region ``II'').  Hence, positive lags
($t_{\gamma} < t_{\rm opt}$) are expected to be produced in regions
``I,'' and negative lags are expected to be produced in regions ``II.''

Since optimal conditions for production of prominent lags are
such that the distance range of enhanced source activity determined by
$x \sim \sigma$ coincides with the region of largest changes of $q$,
one may expect that observed lags, $\Delta t_{\rm lag}$, are produced
at distances $r_m \sim \Delta r_{\rm lag} \simeq c \Delta t_{\rm lag}
{\cal D} \Gamma/(1+z)$.  Hence, when observing a lag of $\Delta t_{\rm
  lag}$, one can estimate $r_m$ and locate the source activity
relative to $r_{\rm BLR}$ and $r_{\rm HDR}$, which values can be
estimated from the broad line luminosities and the hot dust sublimation
temperatures (see \citealt{2009ApJ...704...38S} and
refs.\ therein). Unfortunately,
due to limited effective area of present
$\gamma$-ray instruments (such as \emph{Fermi}-LAT) it is possible to
construct light curves with data quality only sufficient to verify the
presence of correlation between optical and $\gamma$-ray data and
estimate lag length only with time resolution longer than a day (with
exception of the very brightest $\gamma$-ray sources).  Noting that in
luminous blazars $r_{\rm BLR} \sim 10^{18}\,$cm, and that the observed
time of the propagation of the source over the distance range of this
order is $\sim 1.7 (1+z)/(\Gamma/20)^2$ days, presently only lags
produced in the HDR have a chance to be probed observationally.  

To verify the applicability of our lag model,
we use it to match the data
on blazar 3C~279 presented by \cite{Ha12} (see also
\citealt{2010Natur.463..919A}).  A light curve analysis presented in
their paper (R-band vs. $\gamma$-rays of energies $>200\,$MeV) with the
discrete-correlation-function method showed a peak at $+10$
days, indicating a presence of a positive lag of that length.  We used
their data from the time interval MJD~54870-900 including
representative strong flares in optical/$\gamma$-ray bands, and made an
attempt to model them with the flare profiles given in Sec. 2.1.  The data and the model
light curves are presented in Fig.~\ref{fig:3c279fit}.  The parameters
of our model are: $q_{m}=30$, $\beta_{B}=2$, $\beta_{E}=5$, $\sigma=1$,
$r_{m}=4.4\,$pc and $\Gamma=10$.  These results are consistent with the
ERC model parameters used by \cite{Ha12} to fit the broad band spectra
of 3C~279. We would like to emphasize that the presence of a $+10$ days lag in the case of
3C~279 is just a possibility as the used analysis might have been misleaded by poor data quality. Our choice of the strong flares from the whole lightcurve presented in
\cite{Ha12} is therefore just an example indicating that the model can be applied to a real data with very 
reasonable parameters. 

\section{Discussion and conclusions}
\label{sec_disc}

In rarefied magnetized plasmas of astrophysical non-thermal sources,
dissipative processes often generate short-term bursts of relativistic
electrons, which in turn produce synchrotron and inverse-Compton
flares.  Provided that the energy density of seed photons, measured in
the source co-moving frame, is dominated by external sources, the
inverse-Compton radiation is dominated by the ERC process.  For
quasi-uniform energy densities of magnetic and seed radiation fields
across the source, and for electrons energies such that the time scale
of their radiative cooling is much shorter than the time scale of the
burst, the synchrotron and ERC flare profiles will follow the electron
injection function shape almost exactly, without any lags.  At the
same time, the flux ratio of the two spectral components will
approximately be scaled by the ratio of energy densities of seed
radiation field to magnetic field, as long as the inverse Compton
scattering proceeds in the Thomson regime.

\begin{figure}
  \begin{center}
    \includegraphics[width=0.35\textwidth, bb=120 60 493 740, angle=-90]{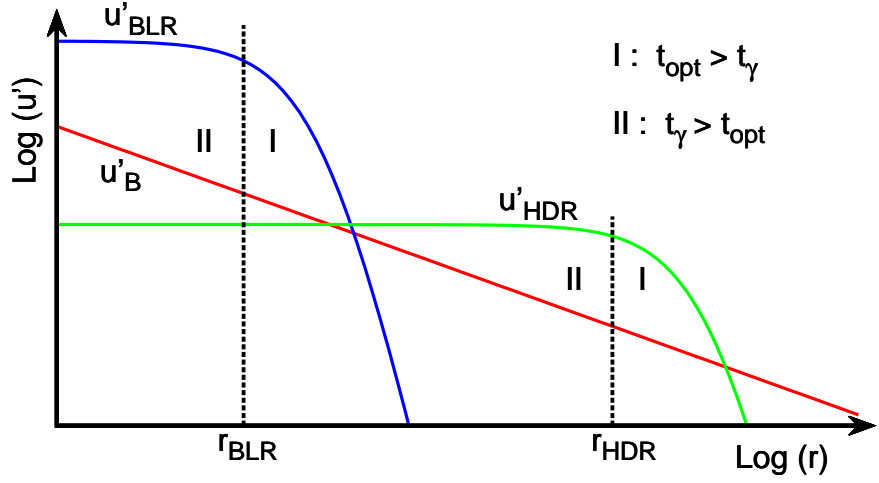}
    \caption{Schematic view of distance dependence of magnetic and
      external radiation field energy densities as measured in the
      source co-moving frame \citep[see Fig.~3 in][]{2009ApJ...704...38S}. Label `I'
      marks the regions where positive lags are expected, whereas label
      `II' corresponds to regions where negative lags are expected.}
   \label{fig:ur}  
 \end{center} 
\end{figure}

\begin{figure}
 \begin{center}
  \includegraphics[width=0.55\textwidth, bb=63 146 552 669, angle=-90]{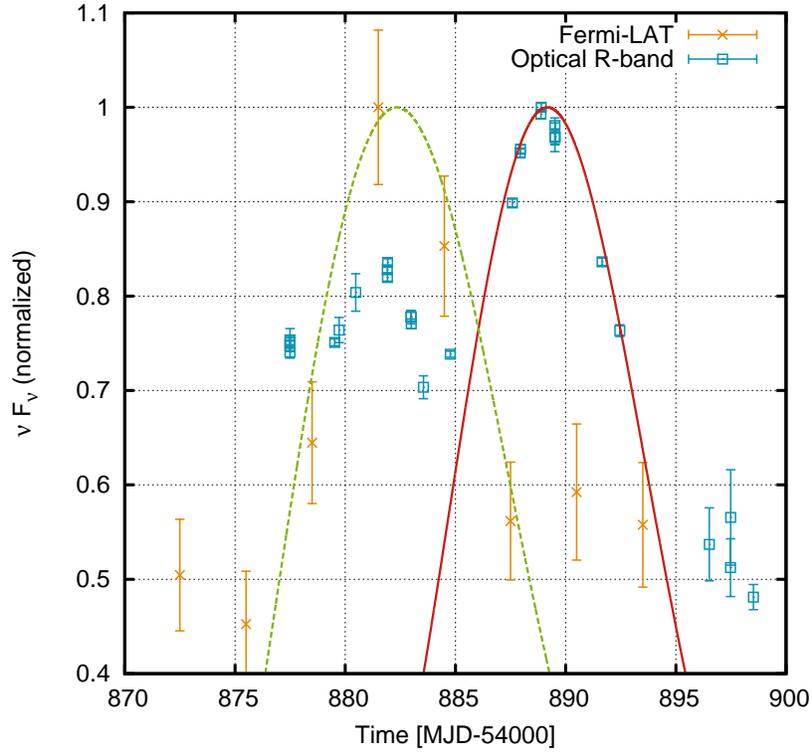}
  \caption{The February 2009 $\gamma$-ray/optical outburst of blazar 3C~279 \citep[from][]{2010Natur.463..919A,Ha12} with a time
    lag of $\approx 10$~days in normalized units.
    Lines show our analytical model
    (\emph{dotted line} --- ERC flare; \emph{solid line} --- synchrotron flare) for parameter values
    $q_{m}=30$, $\beta_{B}=2$, $\beta_{E}=5$, $\sigma=1$,
    $r_{m}=4.4$~pc and $\Gamma=10$.}
  \label{fig:3c279fit}
 \end{center}
\end{figure}

However, in blazars, the outbursts are produced by sources (e.g.,
internal shocks, magnetic reconnection domains) which propagate
down the jet with relativistic speeds.  As we illustrated in
Fig.~\ref{fig:ur}, in FSRQs, the energy density distribution of
external radiation field is significantly stratified, while the magnetic
field intensity is expected to monotonically drop with a distance due
to the jet divergence.  As our equations in Section~\ref{sec_model}
show, in such an environment, for sources propagating down the jet,
optical light curves produced by the synchrotron mechanism differ from
$\gamma$-ray light curves produced by the ERC process.  In particular,
if the source is bursting within the regions marked on
Fig.~\ref{fig:ur} by ``I'' or ``II,'' we will observe flares which are
delayed with respect to each other, with the $\gamma$-ray flares
preceding the optical ones if the burst takes place at $r_{\rm BLR} <
\Delta r_{\rm burst} \ll r_{\rm HDR}$ and $\Delta r_{\rm burst} \gg
r_{\rm HDR}$, and with the optical flares preceding the $\gamma$-ray
ones for $ \Delta r_{\rm burst} < r_{\rm BLR}$ and $r_{\rm BLR} \ll
\Delta r_{\rm burst} < r_{\rm HDR}$.

  Our scheme does not include the presence of a warm dust, which
  together with the hot phase, produces spectra extending up to $20\,\mu$m
  -- $30\,\mu$m (see, e.g., Fig.~7 in \citealt{Sh11}).  The observed
  spectra indicate a distribution of dust extending to much larger
  distances than $r_{\rm HDR}$.  In this case, the decrease of the
  external radiation energy density does not necessarily have to 
  be faster than the
  decrease of the magnetic energy density.  However, frequently observed
  excesses of the NIR radiation over the MIR component extrapolated to
  shorter wavelengths \citep{MNE09,Le10} suggest that there
  can be two separate thermal components from dust, the hot one
  produced by graphite and located in the outer portions of the BLR,
  and the cooler one produced by both graphite and silicate grains and
  enclosed within clumps of the molecular tori extending up to tens of
  parsecs \citep{MN12,Ki12,2012arXiv1205.4543R}.  Hence, a confirmation of 
  predominantly positive optical/$\gamma$-ray lags of a few days can 
  independently support such a stratification of the dust distribution.

Obviously, the model proposed and studied by us, aimed to explain the observed 
lags between optical and $\gamma$-ray flare peaks is not a unique one,
but is the only one which involves {\sl single dissipative events}. 
This model can be verified in future against two-dissipative-zone scenarios,
e.g. those involving reconfinement plus reflection shock structures, or 
two internal shocks formed independently at different locations \citep{kg07}, 
by future more detailed and better resolved light curves.    

\vskip 2em

We studied the dependence of synchrotron-ERC lags on model parameters
which determine the distribution of magnetic and external radiation
fields, their energy density ratio at the time of the electron
injection peak, the dispersion of the injection function, and the
slope of the injected electron spectrum.  Our main results can be
summarized as follows:
\begin{itemize}
\item
Variations of $q=u_{\rm ext}'/u_B'$ with a distance result in
  different ERC and synchrotron light curves;
\item
Electron injection bursts produced within the region of the
  monotonic changes of $q$ lead to the production of the lagged
  flares, positive ($\gamma$-ray first) for ${\rm d}q/{\rm d}r<0$,
  negative for ${\rm d}q/{\rm d}r>0$;
\item
The `prominence' of such lags increases with increasing burst dispersion
  $\sigma$ and increasing gradients of $q$.  The most prominent lags
  are expected when the burst length coincides with the region of
  largest gradient of $q$;
\item
Limited time resolution of $\gamma$-ray data currently 
allows only searches for lags larger than a few days, which corresponds
  to distances of the order of, or larger than, the sublimation radius
  of the hot dust.  As an example,
  we showed that our model light curves can be matched
  to the observed $\gamma$-ray/optical flare of 3C~279
  in February 2009, when a 10-day lag was seen \citep{Ha12}.
\item
Future more detailed and better resolved light curves 
should allow to discriminate 
whether the optical - $\gamma$-ray lags can be explained by single dissipative 
events, or whether two spatially separated dissipation zones must be involved.    

\end{itemize}

\acknowledgments 

We would like to thank the referee Fabrizio Tavecchio for critical comments
which helped to improve the paper.
We acknowledge financial support by
the Polish NCN grant DEC-2011/01/B/ST9/04845,
the NSF grant AST-0907872, the NASA ATP grant NNX09AG02G, and 
NASA Fermi GI grant No. NNX11AO39G.
GM acknowledges DoE support to SLAC via contract No. DE-AE3-76SF00515.


\newpage

\end{document}